\documentclass[prd,twocolumn,nofootinbib,aps,superscriptaddress,tightenlines,preprintnumbers]{revtex4}

\usepackage{epsfig,latexsym,cancel}

\newcommand{\be}{\begin{equation}}
\newcommand{\ee}{\end{equation}}
\newcommand{\bea}{\begin{eqnarray}}
\newcommand{\eea}{\end{eqnarray}}

\newcommand{\mpl}{M_P}

\begin{document}

\preprint{\hbox{CALT-68-2694}  } 
\preprint{\hbox{NPAC-08-05}}

\title{Dark-Matter-Induced Weak Equivalence Principle Violation} 

\author{Sean M. Carroll}
\author{Sonny Mantry}
\affiliation{California Institute of Technology, Pasadena, CA 91125}
\author{Michael J. Ramsey-Musolf}
\affiliation{University of Wisconsin-Madison, Madison, WI 53706}
\affiliation{California Institute of Technology, Pasadena, CA 91125}
\author{Christoper W. Stubbs}
\affiliation{Harvard-Smithsonian Center for Astrophysics, Cambridge, MA 02138}

\begin{abstract}
A long-range fifth force coupled to dark matter can induce a coupling to ordinary
matter if the dark matter interacts with
Standard Model fields.  We consider constraints on such a scenario from both
astrophysical observations and laboratory experiments.   
We also examine the case where the dark matter is a weakly interacting massive
particle, and derive relations between the coupling to dark matter and the coupling
to ordinary matter for different models.  Currently, this scenario is most tightly
constrained by galactic dynamics, but improvements in E\"otv\"os experiments
can probe unconstrained regions of parameter space.
\end{abstract}

\maketitle


A light scalar field $\phi$ 
coupled to dark matter (DM) could mediate a long-range force
of strength comparable to gravity.  A number of models along these lines have been
proposed, motivated both
by attempts to account for features in the distribution of DM and to explore
interactions with quintessence \cite{Damour:1990tw,Friedman:1991dj,
Gradwohl:1992ue,Anderson:1997un,
Carroll:1998zi,Amendola:2001rc,Farrar:2003uw,Gubser:2004du,Gubser:2004uh,
Bertolami:2004nh,Nusser:2004qu,Bean:2007ny}.
Interesting limits on such a force have been derived from observations
of DM dynamics in the tidal stream of the Sagittarius dwarf galaxy 
\cite{Kesden:2006vz,Kesden:2006zb}.
The detection of a new long range force would signal the presence of a new 
mass hierarchy, between the light
scalar mass $m_\phi < 10^{-25}$ eV and the weak scale $m_W \sim 100$ GeV.
Thus, if new long range forces exist in the dark sector, their observation could provide a new window on other puzzling scale hierarchies in particle
physics, such as that between $m_W$ and the Planck scale, $M_P\sim 10^{19}$ GeV.

If the scalar couples to Standard Model (SM) fields, it will give rise to a 
composition-dependent force acting on ordinary matter \cite{Damour:1996xt}; such
forces are tightly constrained by E\"otv\"os experiments looking for violations
of the Weak Equivalence Principle (WEP) \cite{Schlamminger:2007ht}.
On the other hand, even if $\phi$ has \textit{only} an elementary (i.e., renormalizable)
coupling to DM, interactions between DM and ordinary matter will
still induce a coupling of $\phi$ to ordinary matter. This can be thought of as arising from the scalar coupling to 
virtual DM particles
in ordinary atomic nuclei.  We therefore naturally expect a fifth force coupled
to the SM if a light scalar couples to a DM field having SM interactions.  In what 
follows, we show how this scenario may arise in simple model illustrations, and analyze
model-dependent details of its viability in a subsequent publication.  (As this paper
was being prepared for submission, we became aware of closely related work
by Bovy and Farrar \cite{Bovy:2008gh}. For a detailed comparison see ~\cite{Carroll:2009dw}.)

This paper has two goals.  First, we consider the varieties of experimental
constraints on a two-dimensional parameter space, given by the respective
couplings of a new long-range force to ordinary matter and to dark matter.
E\"otv\"os experiments are sensitive to anomalous accelerations
of ordinary matter toward the DM in the galactic center; however, we
argue on general grounds that any force capable of giving rise to a detectable
effect would first give rise to a detectable fifth force acting between two 
sources of ordinary matter.
Our second goal is to explore illustrative scenarios in which a light scalar
couples at tree level only to a Weakly Interacting Massive Particle (WIMP)
DM candidate, and derive the induced coupling to ordinary matter.

\emph{Fifth Force Phenomenology.}
We now turn to order-of-magnitude estimates of the available experimental
constraints on couplings of a light scalar $\phi$ to both ordinary matter and to DM.
We assume the existence of a mechanism that keeps its mass small.  The static potential
between a test object $o$ and a source $s$ arising from the combined effects of gravity
and the coupling to $\phi$ is then
\be
  V = -\frac{GM_oM_s}{r}\left(1 +  \frac{1}{4\pi G}\frac{q_oq_s}{\mu_o\mu_s}\right)\,,
  \label{totalpotential}
\ee
where $q/\mu$ is the charge per unit mass and $G=M_P^{-2}$; for a
fermion $\psi_i$ with mass $m_i$ and  Yukawa coupling
${\mathcal L} = g_i\phi\bar\psi_i \psi_i$, we have $q/\mu = g_i/m_i$.
Searches for WEP-violating fifth forces place limits on the E\"otv\"os parameter,
\be
  \eta = 2\frac{|a_1 - a_2|}{|a_1+a_2|}\,,
  \label{eotvos1}
\ee
where $a_1$ and $a_2$ are the accelerations of two bodies with different compositions.

\begin{figure}
\includegraphics[width=0.45\textwidth]{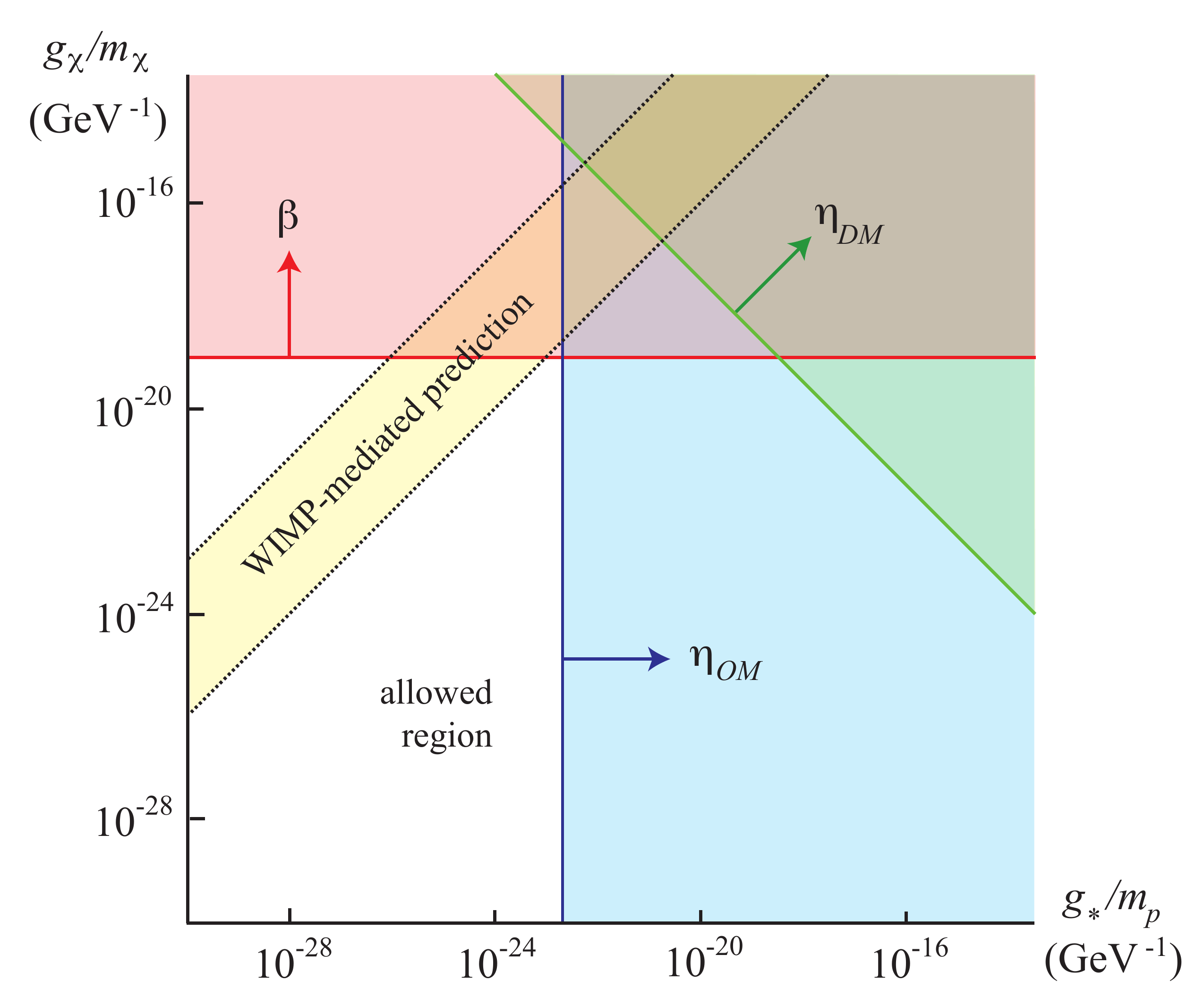}
\caption{Constraints on the strength of fifth forces coupled to ordinary matter
and/or dark matter.  The horizontal axis is the effective coupling to ordinary
matter $g_*$ divided by the proton mass $m_p$, while the vertical axis is the
Yukawa coupling to DM $g_\chi$ divided by the mass $m_\chi$ of the
DM, both in units of inverse GeV.  The vertical blue line is the
constraint from E\"otv\"os experiments with ordinary-matter sources; the horizontal
red line is from DM tidal tails; and the diagonal green line is from
searches for anomalous accelerations in the direction of the Galactic center.
The yellow diagonal band is the range of predictions from the WIMP models
we consider.}
\label{plot}
\end{figure}

We will assume that the dominant couplings of $\phi$ to ordinary matter are
to protons ($p$), neutrons ($n$), and electrons ($e$), neglecting, for example,
couplings to atomic binding energy.  
(See {\it e.g.} \cite{Kaplan:2000hh,Dvali:2001dd,Dent:2008gu}.) 
Then for a neutral object made of ordinary matter we can define
\be
 \frac{g_p + g_e}{m_p + m_e} \equiv  \frac{g_*}{m_p}  \,,
  \quad \frac{g_n}{m_n}\equiv  (1+\epsilon) \frac{g_*}{m_p}\,.  
  \label{gstar}
\ee
The parameter $\epsilon$ can be calculated by first computing the coupling of $\phi$
to nucleons by matching onto its coupling to quarks,
$g_N = \sum g_i \langle N|\bar{q}_i q_i|N\rangle$,
where the sum is over all quarks.  We also assume that $g_i$ for
quarks and leptons is proportional to the mass of the fermion, so we can write
$g_i = \bar{g}m_i/m_p$; this assumption holds true in simple
models, as we show below.  
We obtain $g_N$ by comparing the energy momentum tensor in six-flavor
and three-flavor QCD \cite{Shifman:1978zn} and using the known light-quark 
nucleon matrix elements \cite{Belanger:2008sj}.
We find $g_p\sim 0.4 \bar g \approx g_n$.
For the electron we have $g_e = \bar g m_e/m_p$, which leaves us
with $\epsilon=-8\times 10^{-4}$.

Assuming that the fifth force is much weaker than gravity,
the E\"otv\"os parameter with both test bodies and source constructed from 
ordinary matter is then
\be
  \eta_{_{OM}} \approx \frac{\mpl^2}{4\pi} \left|\epsilon (f_{1}-f_2)\right|
  \left(\frac{g_*}{m_p}\right)^2\,,
\ee
where $f_i = Z_i/A_i$ is the nuclear fraction of protons in body $i$, and 
``$OM$'' denotes a source consisting of ordinary matter. Atomic binding energy 
corrections may introduce $\mathcal{O}(1)$ corrections to this relation but do not 
substantially affect our conclusions\cite{Kaplan:2000hh}. 
The current best limit on $\eta$ from torsion-balance experiments involving 
terrestrial test bodies is $\eta_{_{OM}} < 2\times10^{-13}$, using materials 
with $f_1-f_2 \sim 2\times 10^{-2}$ \cite{Schlamminger:2007ht}.  From this we derive
$g_*/m_p <  3\times 10^{-23}$~GeV$^{-1}$, which we have plotted as a vertical line
in Fig.~\ref{plot}.

Now consider a fermionic dark matter particle $\chi$ coupled via
${\mathcal L} = g_\chi \phi \bar\chi\chi$.  Interesting limits may be obtained on 
anomalous accelerations of laboratory test bodies in the direction of the 
galactic center, where the dominant source is presumably the DM
inside the Solar circle \cite{Stubbs:1993xk}.
For a source made of DM and test bodies of ordinary matter, we obtain
\be
  \eta_{_{DM}} \approx \frac{\mpl^2}{4\pi} \left|\epsilon (f_{1}-f_2)\right|
  \frac{g_*g_\chi}{m_pm_\chi}\,,
\ee
where ``$DM$'' denotes accelerations sourced by dark matter.  
The best current limits on anomalous accelerations in the direction of the
galactic center give $\eta_{_{DM}} < 10^{-5}$ \cite{Schlamminger:2007ht},
corresponding to $g_*g_\chi/m_pm_\chi < 5\times 10^{-38}$~GeV$^{-2}$.
This is plotted as a downward-sloping diagonal line in Fig.~\ref{plot}.

Separate limits on $g_\chi/m_\chi$ are obtained from astrophysical tests, such
as the dynamics of galactic tidal streams \cite{Kesden:2006vz,Kesden:2006zb}.
In this case, the constraint limits the strength of the force due to $\phi$ relative
to that due to gravity, rather than a composition-dependent acceleration.  The 
relevant parameter is 
\be
  \beta = \frac{\mpl}{\sqrt{4\pi}}\frac{g_\chi}{m_\chi}\,.
\ee
For reasonable models of the Sagittarius tidal stream, we obtain $\beta < 0.2$ 
\cite{Kesden:2006vz,Kesden:2006zb}, corresponding to 
$g_\chi/m_\chi < 10^{-19}$~GeV$^{-1}$, 
plotted as a horizontal line in Fig.~\ref{plot}.

From Fig.~\ref{plot}, it is clear that current bounds from constraints on 
anomalous accelerations toward the galactic center do not cover any region of 
parameter space that is not already excluded by constraints on the couplings
to ordinary matter and DM alone.  
Assuming that limits on $\beta$  do not appreciably
improve,  $\eta_{_{DM}}$ will only probe unconstrained  parameter
space once it is  more sensitive to $g_*/m_p$ 
for $g_\chi/m_\chi \sim 10^{-19}$~GeV$^{-1}$ (the value along the
the $\beta$ constraint line) than $\eta_{_{OM}}$. 
Although any improvement in sensitivity to $|a_1-a_2|$ will lead to the same
improvements in $\eta_{_{DM}}$ and $\eta_{_{OM}}$,   they depend linearly 
and quadratically on $g_*/m_p$, respectively. 
Since  the $\eta_{_{DM}}$ constraint  currently is weaker by a factor of $10^{4}$ than the
$\eta_{_{OM}}$ bound along the $\beta$ constraint line, one must improve sensitivity 
to $|a_1-a_2|$ by at least $\sim10^8$.
 This corresponds to $\eta_{_{OM}} \sim 10^{-21}$, $\eta_{_{DM}} \sim 10^{-13}$, $g_*/m_p \sim 10^{-27}$~GeV$^{-1}$.
  The proposed STEP experiment aims at $\eta_{_{OM}} < 10^{-17}$ \cite{1993STIN...9526184B}, not enough
to achieve this goal.
If an anomalous acceleration toward the galactic center were detected with
$\eta_{_{DM}} > 10^{-13}$ but with no corresponding detection of $\eta_{_{OM}}$, 
it could not be accommodated by the type of theory considered here.

\emph{Model Examples: WIMP Dark Matter.}
The most popular DM models involve WIMPs -- stable neutral particles $\chi$ living in 
some representation
of the electroweak gauge and Poincar\'e groups.  Their interactions with 
electroweak gauge bosons 
provide an annihilation cross-section that leads to cosmologically interesting
relic abundances.  A classification of the various possible representations 
containing a viable DM candidate can be found in \cite{Cirelli:2005uq}.
Here, we select a few cases 
to illustrate the range of scenarios for WIMP-induced
fifth-force couplings to ordinary matter.

\begin{figure}
\includegraphics[width=0.3\textwidth]{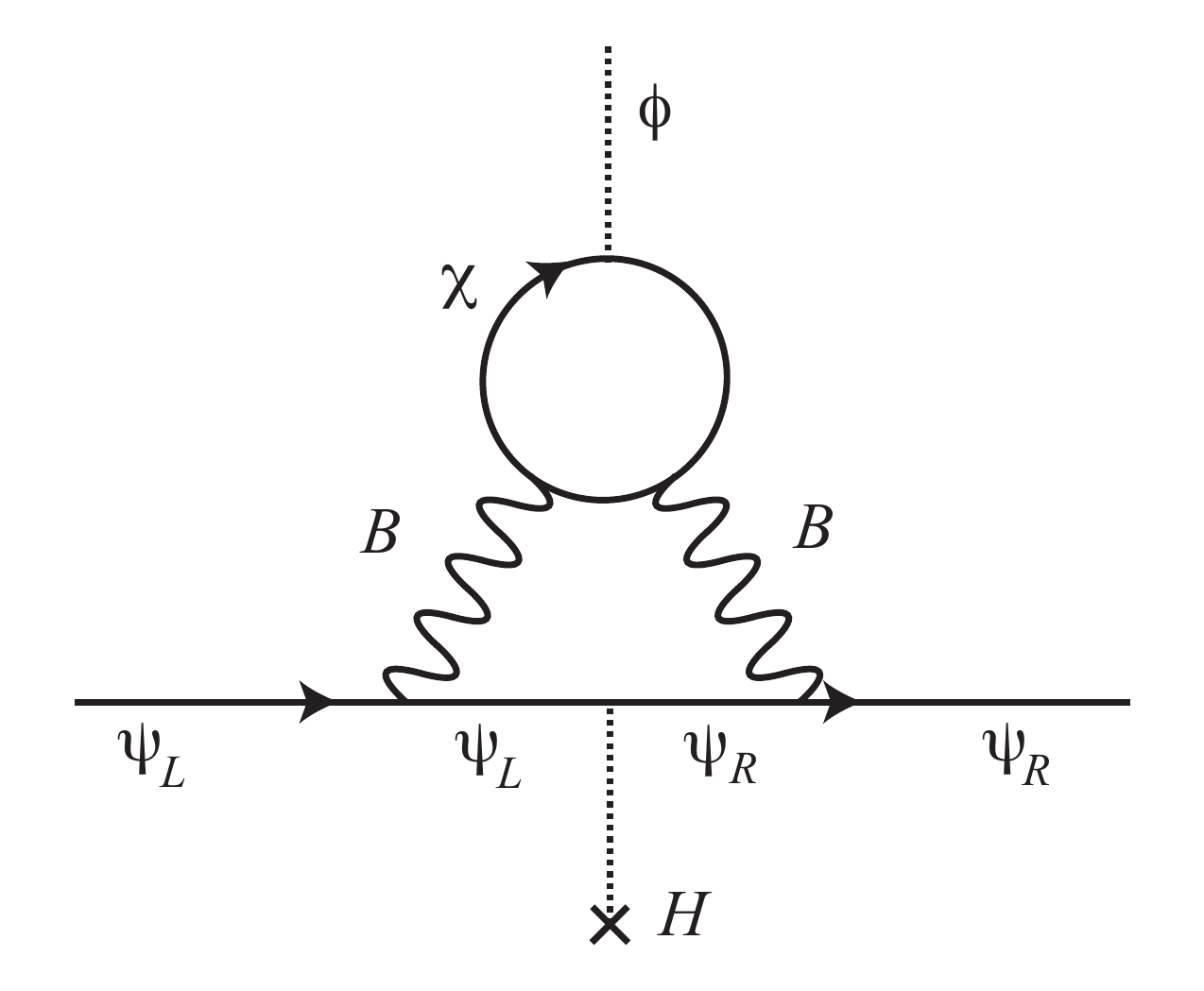}
\caption{Two-loop graph inducing an interaction between a massless scalar $\phi$
and a Dirac fermion $\psi$, mediated by a WIMP $\chi$ and hypercharge gauge
bosons $B$.}
\label{2loop}
\end{figure}

\begin{figure}
\includegraphics[width=0.35\textwidth]{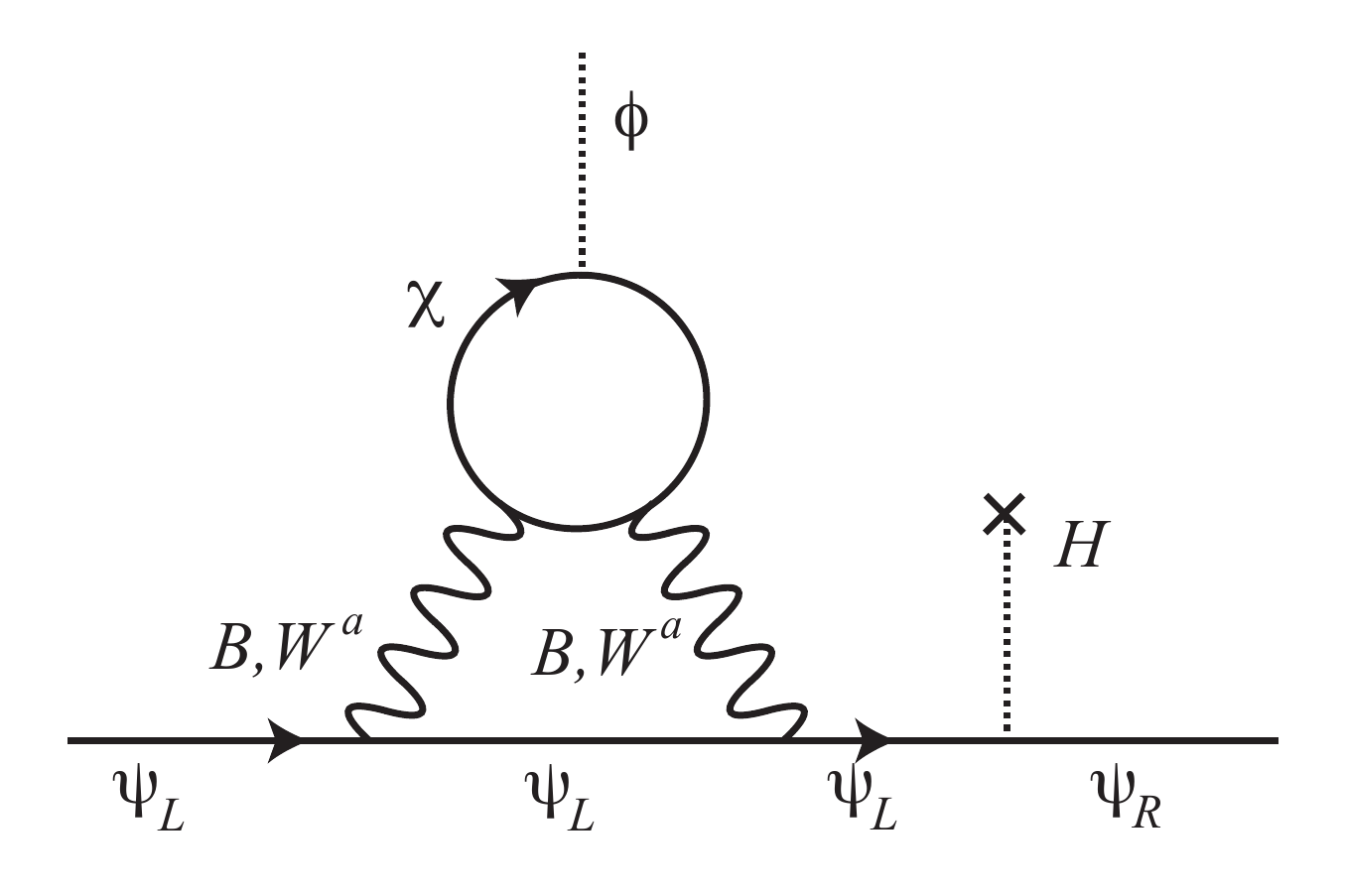}
\caption{Two-loop \lq\lq Compton scattering"  graph inducing an interaction between a massless scalar $\phi$
and a Dirac fermion $\psi$, mediated by a WIMP $\chi$ and electroweak gauge bosons.}
\label{3loop}
\end{figure}

Figs.~\ref{2loop} and \ref{3loop} show the lowest-order processes that generate 
the gauge-invariant interaction 
\be
\label{eq:yukawa}
c_i \phi\bar{\psi}_{iL } H \psi_{iR }/\Lambda+\mathrm{h.c.}
\ee
involving $\phi$, 
the Higgs doublet $H$ and the
left-handed doublet and right-handed singlet components of a SM fermion $\psi_i$. 
Here,  $\Lambda$ is a mass scale associated with the particles in the loops and $c_i$ 
is proportional to the fermion Yukawa coupling $y_i$. 
After electroweak symmetry breaking, in which the neutral component of $H$ obtains a 
vacuum expectation value $v = 246$~GeV, this interaction yields the coupling 
$g_i\phi\bar{\psi}_i\psi_i$ with $g_i\sim m_i/\Lambda$. 

Since right-handed fermions do not couple to
the SU(2)$_L$ gauge bosons $W^a$, Fig.~\ref{2loop} only contributes
if the WIMP $\chi$ has nonvanishing hypercharge.  In that case, the hypercharge
gauge boson $B$, can couple $\chi$ to both $\psi_{iL}$ and $\psi_{iR}$. The \lq\lq Compton scattering" process of Fig.~\ref{3loop}, in contrast, contributes for all WIMPs, with those having $Y=0$ receiving contributions only from internal SU(2)$_L$ gauge bosons. The two-loop subgraph of Fig.~\ref{3loop} generates the structure $\cancel{P}+\cancel{P}^{\, \prime}$ involving the incoming and outgoing fermion momenta. The $\cancel{P}^{\, \prime}$ cancels the $1/\cancel{P}^{\, \prime}$ of the intermediate $\psi_L$, leading to the momentum-independent interaction (\ref{eq:yukawa}). Formally, the two-loop subgraph of Fig.~\ref{3loop} -- along with diagrams involving external gauge boson insertions (not shown) -- yields the operator ${\bar\psi}_L i(\overleftarrow{\cancel{D}}-\overrightarrow{\cancel{D}})\psi_L$ that is equivalent to (\ref{eq:yukawa}) by virtue of the equation of motion for $\psi_L$. 

The simplest realization of this scenario occurs when $\chi$ is a scalar 
SU(2)$_L$ doublet, having an elementary $g_\chi\phi \chi^\dag\chi$ interaction 
with the long-range force mediator. A mass $m_\chi\sim 0.5$~TeV is needed to obtain 
the observed DM relic density \cite{Cirelli:2005uq}. 
A fermionic realization requires the presence of two 
doublets with $Y=\pm 1$ to cancel anomalies and $m_\chi$ of order 1~TeV. For either 
case, the graph in Fig.~\ref{2loop} is finite and a simple estimate yields
\be
  g_i \sim \left(\frac{\alpha_{\rm em}}{4\pi}\right)^2 \frac{m_i}{m_\chi}g_\chi\,.
  \label{2loophyper}
\ee
The contribution from Fig.~\ref{3loop} is na\"ively an order of magnitude larger since it involves four powers of the SU(2)$_L$ gauge coupling that is roughly twice as large as the U(1)$_Y$ coupling entering Fig.~\ref{2loop}:
\be
  g_i \sim \left(\frac{\alpha_{\rm em}}{\pi}\right)^2 \frac{m_i}{m_\chi}g_\chi\,.
    \label{2loopw}
\ee
In terms of the nucleon coupling $g_*$ defined in (\ref{gstar}), the estimate (\ref{2loophyper}) becomes
\be
\label{eq:gstar}
  \frac{g_*}{m_p} \sim 10^{-7}\frac{g_\chi}{m_\chi}\,.
\ee
This provides a lower limit (upper dotted line in Fig.~\ref{plot}) on the coupling strength induced by WIMP dark matter,
and would be relevant if $\chi$ were a singlet of $SU(2)_L$ with
non-zero hypercharge.  If $\chi$ is a doublet or triplet of $SU(2)_L$,
equation (9) applies, and the resulting coupling is an order of magnitude larger, 
$g_*/m_p \sim 10^{-6} g_\chi/m_\chi$.

\begin{figure}
\includegraphics[width=0.35\textwidth]{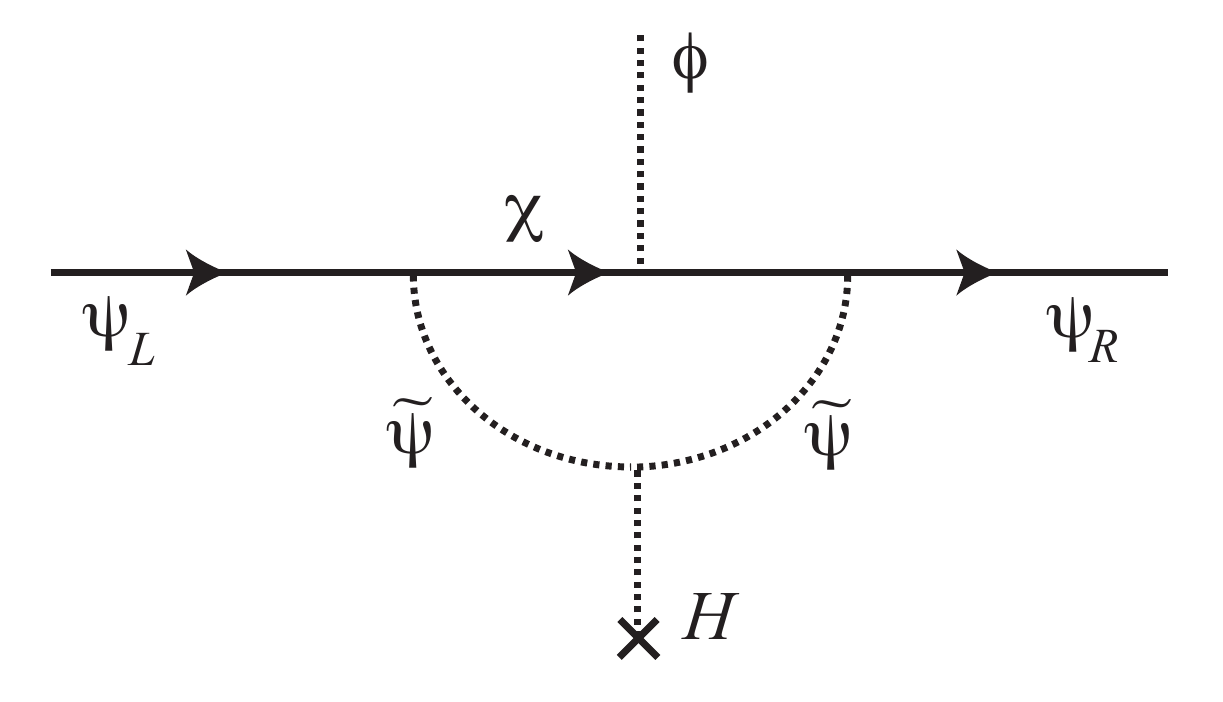}
\caption{One-loop graph involving scalar sfermions $\tilde\psi$.}
\label{1loop}
\end{figure}

Realistic implementations of the WIMP idea often introduce more than just a single 
DM field (and its charged partners). In supersymmetry, for example, the $\chi$ is the 
lightest supersymmetric particle (LSP) and is a linear superposition of the superpartners 
of the electroweak gauge bosons (winos and binos) and Higgs bosons (higgsinos). In 
addition, one has scalar squarks and sleptons $\tilde\psi$ that interact with their partners 
and the LSP via an interaction $\lambda{\tilde\psi}{\bar\psi}\chi+\mathrm{h.c.}$ If $\phi$ is 
the scalar component of a singlet superfield ${\hat S}$, a superpotential term of the form 
${\hat S} {\hat H}_u\cdot{\hat H}_d$ will generate a coupling of $\phi$ to the higgsino 
components of the LSP. As a result, we expect a one-loop coupling of $\phi$ to SM 
fermions, as shown in Fig.~\ref{1loop}, which gives
\be
  g_i \sim \frac{1}{16\pi^2} \frac{m_i \mu \lambda^2}{M_{\rm susy}^2 }g_\chi\,,
\ee
where $\mu$ is the
$\mu$-parameter of supersymmetry and $M_{\rm susy}$ is the mass of the heaviest
superparticle in the loop. We assume for simplicity that $m_\chi \approx M_{\rm susy} \approx
\mu \approx v$.  The most favorable case is when 
the DM $\chi$ is primarily a bino, in which case 
$\lambda = g_1$, the U(1)$_Y$ gauge coupling.  This implies
\be
  \frac{g_*}{m_p} \sim 10^{-4}\frac{g_\chi}{m_\chi}\,,
\ee
leading to the lower dotted line in Fig.~\ref{plot}.
In this case, where the existence of the squarks enables a one-loop contribution,
the coupling to ordinary matter is enhanced by a factor of order $10^3$.  This is the
most optimistic scenario, in the sense of creating the strongest coupling of $\phi$
to ordinary matter.  If $\chi$ is predominantly higgsino, the coupling $\lambda$ will
be approximately the Higgs Yukawa coupling to $\psi_i$, which is order one for the
top quark. 
Fig.~\ref{plot} plots the range of values in different models,
from a minimal WIMP scenario that only couples $\phi$ to ordinary matter at
two loops, to a bino-like model (with sfermions) that implies a
contribution at one loop.  Higgsino-like DM, to be considered in future work, could live outside the WIMP band in 
Fig.~\ref{plot} due to a Yukawa  enhanced coupling to the top quark. 
 For $\phi$ coupling to ordinary
matter by mixing with the Higgs, see ~\cite{Carroll:2009dw}.
Similarly, in any given model of non-sterile DM coupled to a long range force, 
one can determine the size of the  induced coupling to ordinary matter and
its implications for E\"otv\"os experiments.

\emph{Conclusions.}
We have found that a weakly-interacting dark matter particle $\chi$ coupled
to a light scalar $\phi$ with strength $g_\chi$ naturally induces an effective
coupling to ordinary matter with strength
${g_*}/{m_p} \sim (10^{-7} - 10^{-4}) {g_\chi}/{m_\chi}$.
The low end of this range corresponds to minimal WIMP models with only
higher-loop contributions to the interaction of $\phi$ with SM fields,
through SU(2)$_L$ gauge bosons, while the high end
corresponds to bino-like DM with one-loop contributions.
Comparing with Fig.~\ref{plot}, we see that the best current limits
on these models come from purely astrophysical bounds on new long-range
forces in the dark sector; if improvements in these techniques discovered such
a force, it would predict a new force between ordinary matter if the DM
were WIMPs.  Meanwhile, improvements in the sensitivity
of E\"otv\"os-type experiments could provide interesting new constraints on
a WIMP-mediated coupling.
Any improvement of the limits on $g_*/m_p$ 
would begin cutting into the predictions of the models examined here
(at the order-of-magnitude precision we considered).  
Currently, constraints on a fifth force in the direction of the
galactic center do not independently constrain any of the parameter space;
measurements of anomalous accelerations would
have to improve by a factor of about $10^8$ before they would begin to do so.

We thank J. Distler, M. Kamionkowski, 
T. Tait, M. Wise, and K. Zurek for helpful 
conversations.  This research was supported in part by 
Department of Energy contracts
DE-FG03-92-ER40701 and  DE-FG02-08ER41531, the Wisconsin Alumni Research 
Foundation, and the Gordon and Betty Moore Foundation.

\end{document}